\documentstyle{lamuphys}
\begin{document}
\title{ Vertex Operators and  Solitons of
Constrained KP Hierarchies }
\author{H. Aratyn\inst{1}, L.A. Ferreira\inst{2}, J.F. Gomes\inst{2} and 
A.H. Zimerman\inst{2}}
\institute{Department of Physics,
University of Illinois at Chicago,
845 W. Taylor St., Chicago, IL 60607-7059
\and Instituto de F\'\i sica Te\'orica - IFT/UNESP,
Rua Pamplona 145, 
01405-900, S\~ao Paulo - SP, Brazil}

\maketitle 

\begin{abstract}
We construct the vertex operator representation for the Affine Kac-Moody
${\hat{sl}}(M+K+1)$ algebra, which is relevant for the construction of the
soliton solutions of the constrained KP hierarchies. The oscillators 
involved in the vertex operator construction are provided by the 
Heisenberg subalgebras of ${\hat{sl}}(M+K+1)$ realized in 
the unconventional gradations.
The well-known limiting cases are   
the homogeneous Heisenberg subalgebra of ${\hat{sl}}(M+1)$ and the principal 
Heisenberg subalgebra of ${\hat{sl}}(K+1)$. 
The explicit example of $M=K=1$ is discussed in detail and
the corresponding soliton solutions and tau-functions are given. 

\end{abstract}

\def\cKP{{\sf cKP}~}
\def\bfs{{\bf s}}
\def\bfts{{\bf {\tilde s}}}
\def\qs{Q_{\bfs}}
\def\kere{\mbox{\rm Ker (ad $E$)}}
\def\ime{\mbox{\rm Im (ad $E$)}}
\def\cgh{{\widehat {\cal G}}}
\def\vp{\varphi}
\def\qsp{Q_{\bfs^{\pr}}}
\def\bfsp{{\bf s}^{\prime}}
\newcommand\ket[1]{\vert {#1}\rangle}
\def\bra#1{\langle #1 \mid}


\def\rf#1{(\ref{eq:#1})}
\def\lab#1{\label{eq:#1}}
\def\nonu{\nonumber}
\def\br{\begin{eqnarray}}
\def\er{\end{eqnarray}}
\def\be{\begin{equation}}
\def\ee{\end{equation}}
\def\eq{\!\!\!\! &=& \!\!\!\! }
\def\foot#1{\footnotemark\footnotetext{#1}}
\def\lb{\lbrack}
\def\rb{\rbrack}
\def\llangle{\left\langle}
\def\rrangle{\right\rangle}
\def\blangle{\Bigl\langle}
\def\brangle{\Bigr\rangle}
\def\llb{\left\lbrack}
\def\rrb{\right\rbrack}
\def\Blb{\Bigl\lbrack}
\def\Brb{\Bigr\rbrack}
\def\lcurl{\left\{}
\def\rcurl{\right\}}
\def\({\left(}
\def\){\right)}
\def\v{\vert}                     
\def\bv{\bigm\vert}               
\def\Bgv{\;\Bigg\vert}            
\def\bgv{\bigg\vert}              
\def\lskip{\vskip\baselineskip\vskip-\parskip\noindent}
\def\mskp{\par\vskip 0.3cm \par\noindent}
\def\sskp{\par\vskip 0.15cm \par\noindent}
\def\bc{\begin{center}}
\def\ec{\end{center}}
\def\Lbf#1{{\Large {\bf {#1}}}}
\def\lbf#1{{\large {\bf {#1}}}}

\newcommand{\sect}[1]{\setcounter{equation}{0}\section{#1}}
\renewcommand{\theequation}{\thesection.\arabic{equation}}
\relax


\def\tr{\mathop{\rm tr}}                  
\def\Tr{\mathop{\rm Tr}}                  
\newcommand\partder[2]{{{\partial {#1}}\over{\partial {#2}}}}
\newcommand\partderd[2]{{{\partial^2 {#1}}\over{{\partial {#2}}^2}}}
\newcommand\partderh[3]{{{\partial^{#3} {#1}}\over{{\partial {#2}}^{#3} }}}
\newcommand\partderm[3]{{{\partial^2 {#1}}\over{\partial {#2} \partial {#3} }}}
\newcommand\partderM[6]{{{\partial^{#2} {#1}}\over{{\partial {#3}}^{#4}
{\partial {#5}}^{#6} }}}          
\newcommand\funcder[2]{{{\delta {#1}}\over{\delta {#2}}}}
\newcommand\Bil[2]{\Bigl\langle {#1} \Bigg\vert {#2} \Bigr\rangle}  
\newcommand\bil[2]{\left\langle {#1} \bigg\vert {#2} \right\rangle} 
\newcommand\me[2]{\left\langle {#1}\right|\left. {#2} \right\rangle} 

\newcommand\sbr[2]{\left\lbrack\,{#1}\, ,\,{#2}\,\right\rbrack} 
\newcommand\Sbr[2]{\Bigl\lbrack\,{#1}\, ,\,{#2}\,\Bigr\rbrack} 
\newcommand\pbr[2]{\{\,{#1}\, ,\,{#2}\,\}}       
\newcommand\Pbr[2]{\Bigl\{ \,{#1}\, ,\,{#2}\,\Bigr\}}  
\newcommand\pbbr[2]{\lcurl\,{#1}\, ,\,{#2}\,\rcurl}  


\def\a{\alpha}
\def\b{\beta}
\def\c{\chi}
\def\d{\delta}
\def\D{\Delta}
\def\eps{\epsilon}
\def\vareps{\varepsilon}
\def\g{\gamma}
\def\G{\Gamma}
\def\grad{\nabla}
\def\h{{1\over 2}}
\def\l{\lambda}
\def\L{\Lambda}
\def\m{\mu}
\def\n{\nu}
\def\o{\over}
\def\om{\omega}
\def\Om{\Omega}
\def\p{\phi}
\def\P{\Phi}
\def\pa{\partial}
\def\tpa{{\tilde \partial}}
\def\pr{\prime}
\def\ra{\rightarrow}
\def\lra{\longrightarrow}
\def\s{\sigma}
\def\S{\Sigma}
\def\t{\tau}
\def\th{\theta}
\def\Th{\Theta}
\def\z{\zeta}
\def\ti{\tilde}
\def\wti{\widetilde}
\newcommand\sumi[1]{\sum_{#1}^{\infty}}   
\newcommand\fourmat[4]{\left(\begin{array}{cc}  
{#1} & {#2} \\ {#3} & {#4} \end{array} \right)}
\newcommand\twocol[2]{\left(\begin{array}{cc}  
{#1} \\ {#2} \end{array} \right)}


\def\cA{{\cal A}}
\def\cB{{\cal B}}
\def\cC{{\cal C}}
\def\cD{{\cal D}}
\def\cE{{\cal E}}
\def\cF{{\cal F}}
\def\cG{{\cal G}}
\def\cH{{\cal H}}
\def\cI{{\cal I}}
\def\cJ{{\cal J}}
\def\cK{{\cal K}}
\def\cL{{\cal L}}
\def\cM{{\cal M}}
\def\cN{{\cal N}}
\def\cO{{\cal O}}
\def\cP{{\cal P}}
\def\cQ{{\cal Q}}
\def\cR{{\cal R}}
\def\cS{{\cal S}}
\def\cT{{\cal T}}
\def\cU{{\cal U}}
\def\cV{{\cal V}}
\def\cW{{\cal W}}
\def\cY{{\cal Y}}
\def\cZ{{\cal Z}}

\def\phanta{\phantom{aaaaaaaaaaaaaaa}}
\def\phantb{\phantom{aaaaaaaaaaaaaaaaaaaaaaaaa}}
\def\phantc{\phantom{aaaaaaaaaaaaaaaaaaaaaaaaaaaaaaaaaaa}}


\def\lie{{\cal G}}
\def\dlie{{\cal G}^{\ast}}
\def\elie{{\widetilde \lie}}
\def\edlie{{\elie}^{\ast}}
\def\hlie{{\cal H}}
\def\wlie{{\widetilde \lie}}
\def\ulie{{\cal U}\( {\cal G}\)}           
\def\f#1#2#3{f^{{#1}{#2}}_{#3}}            


\font\numbers=cmss12
\font\upright=cmu10 scaled\magstep1
\def\stroke{\vrule height8pt width0.4pt depth-0.1pt}
\def\topfleck{\vrule height8pt width0.5pt depth-5.9pt}
\def\botfleck{\vrule height2pt width0.5pt depth0.1pt}
\def\Zmath{\vcenter{\hbox{\numbers\rlap{\rlap{Z}\kern 0.8pt\topfleck}\kern
2.2pt
                   \rlap Z\kern 6pt\botfleck\kern 1pt}}}
\def\Qmath{\vcenter{\hbox{\upright\rlap{\rlap{Q}\kern
                   3.8pt\stroke}\phantom{Q}}}}
\def\Nmath{\vcenter{\hbox{\upright\rlap{I}\kern 1.7pt N}}}
\def\Cmath{\vcenter{\hbox{\upright\rlap{\rlap{C}\kern
                   3.8pt\stroke}\phantom{C}}}}
\def\Rmath{\vcenter{\hbox{\upright\rlap{I}\kern 1.7pt R}}}
\def\IZ{\ifmmode\Zmath\else$\Zmath$\fi}
\def\IQ{\ifmmode\Qmath\else$\Qmath$\fi}
\def\IN{\ifmmode\Nmath\else$\Nmath$\fi}
\def\IC{\ifmmode\Cmath\else$\Cmath$\fi}
\def\IR{\ifmmode\Rmath\else$\Rmath$\fi}

\def\one{\hbox{{1}\kern-.25em\hbox{l}}}
\def\0#1{\relax\ifmmode\mathaccent"7017{#1}%
        \else\accent23#1\relax\fi}
\def\omz{\0 \omega}


\def\ltimes{\mathrel{\vrule height1ex}\joinrel\mathrel\times}
\def\rtimes{\mathrel\times\joinrel\mathrel{\vrule height1ex}}


\def\mark{\noindent{\bf Remark.}\quad}
\def\prop{\noindent{\bf Proposition.}\quad}
\def\exam{\noindent{\bf Example.}\quad}

\newtheorem{definition}{Definition}[section]
\newtheorem{proposition}{Proposition}[section]
\newtheorem{theorem}{Theorem}[section]
\newtheorem{lemma}{Lemma}[section]
\newtheorem{corollary}{Corollary}[section]
\def\proof{\par{\it Proof}. \ignorespaces} \def\endproof{{$\Box$}\par}
\newenvironment{Proof}{\proof}{\endproof}



\def\Ouc{{\cal O}_{(U_0 ,c)}}           
\def\Gsu{G_{stat}(U_0 ,c)}                   
\def\Gs{G_{stat}}                            
\def\Asu{{\lie}_{stat} (U_0 ,c)}                   
\def\As{{\lie}_{stat}}                             
\def\Suc#1{\Sigma \Bigl( #1 ; (U_0 ,c) \Bigr)}        
\def\suc#1{{\hat \sigma} (#1 ; (U_0 ,c))}
\def\sh{\hat s}                              
\def\ssh#1{{\hat \sigma}^{#1}}
\def\Y#1{Y(#1)}
\def\y{{\hat y}}
\def\yp{y_{+}(g^{-1})}
\def\YT{Y_t (g^{-1})}
\def\yt{Y_t (g)}
\def\W1#1{W \lbrack #1 \rbrack}                
\def\Wuc#1{W \lbrack #1 ; (U_0 ,c)\rbrack}    


\def\hd{{\widehat D}}
\def\dt{{\hat d}}
\def\Gpr{G^{\pr}}
\def\GT{\tilde \Gamma}
\def\GTy{\funcder {\GT} {y (t)}}
\def\GTz#1{\funcder {\GT} {y_{#1} (t)}}
\def\Ly#1{{\hat L}^{#1}_t (y)}
\def\Ry#1{R^{#1}_t (y)}
\def\LA{{\cal L}^A}
\def\ME#1#2{\left\langle #1\right|\left. #2 \right\rangle}
\def\contract{\makebox[1.2em][c]{
        \mbox{\rule{.6em}{.01truein}\rule{.01truein}{.6em}}}\,}
\def\xf{{\rm Ham} (f)}
\def\dc{{\cal D}}
\def\tg{T^{\ast} G}


\def\Tu#1{{\widetilde \Theta}^{#1}}
\def\Td#1{{\widetilde \Theta}_{#1}}
\def\Z{\widetilde Z}
\def\T#1{{\hat {\cal T}}(#1)}
\def\dNz{\delta^{(N)} (z_1 - z_2 )}
\def\dNth{\delta^{(N)} (\th_1 - \th_2 )}
\def\dN#1#2{\delta^{(N)} ({#1} - {#2})}
\def\DN{{\llbrack D {\widetilde \Theta} \rrbrack}^2_N}
\def\Du#1{{\widetilde D}^{#1}}
\def\Dd#1{{\widetilde D}_{#1}}


\def\Tor{{\wti {\rm SDiff}}\, (T^2 )}
\def\Lh#1{{\hat {\cal L}}({#1})}
\def\M{{\cal M}}
\def\dM{{\cal M}^{\ast}}
\def\Mc{{\cal M}(R^1 \times S^1 )}
\def\dMc{{\cal M}^{\ast}(R^1 \times S^1 )}
\def\st1{\stackrel{\ast}{,}}

\def\symp#1{{\cal S}{\cal D}if\!\! f \, ({#1})}
\def\esymp#1{{\wti {\cal S}{\cal D}if\!\! f} \, ({#1})}
\def\Symp#1{{\rm SDiff}\, ({#1})}
\def\eSymp#1{{\wti {\rm SDiff}}\, ({#1})}
\def\vol#1{{{\cal D}if\!\! f}_0 ({#1})}
\def\Vol#1{{\rm Diff}_0 ({#1})}


\def\Winf{{\bf W_\infty}}               
\def\Win1{{\bf W_{1+\infty}}}           
\def\nWinf{{\bf {\hat W}_\infty}}       
\def\Winft#1{{\bf W_\infty^{\geq {#1}}}}    
\def\winf{{\bf w_\infty}}
\def\win1{{\bf w_{1+\infty}}}
\def\hWinf{{\bf {\hat W}_{\infty}}}        
\def\DO{DOP (S^1 )}                           
\def\eDOS{{\widetilde {DOP}} (S^1 )}          
\def\eDO {{\widetilde {DOP}}}                 
\def\DA{{\cal DOP} (S^1 )}                    
\def\eDAS{{\widetilde {\cal DOP}} (S^1 )}     
\def\eDA{{\widetilde {\cal DOP}}}     
\def\dDA{{\cal DOP}^{\ast} (S^1 )}                  
\def\edDAS{{\widetilde {\cal DOP}}^{\ast} (S^1 )}   
\def\edDA{{\widetilde {\cal DOP}}^{\ast}}   
\def\DOP#1{{DOP (S^1 )}_{\geq{#1}}}                 
\def\DAP#1{{{\cal DOP} (S^1 )}_{\geq{#1}}}          
\def\eDAP#1{{{\widetilde {\cal DOP}} (S^1 )}_{\geq{#1}}}
\def\dDAP#1{{{\cal DOP}^{\ast} (S^1 )}_{\geq{#1}}}     
\def\edDAP#1{{{\widetilde {\cal DOP}}^{\ast} (S^1 )}_{\geq{#1}}}
\def\stc{\stackrel{\circ}{,}}              
\def\sto{\stackrel{\otimes}{,}}              
\def\sta{\, ,\,}
\def\xx{(\xi , x)}
\def\yy{(\zeta , y)}
\def\xxt{(\xi , x ; t )}
\def\intres{\int dx\, {\rm Res}_\xi \; }
\def\Intres{\int dx\, {\rm Res} \; }
\def\intrest{\int dt\, dx\, {\rm Res}_\xi \;}
\def\Intrest{\int dt\, dx\, {\rm Res} \;}
\def\Res{{\rm Res}_\xi}
\def\pexx{e^{\pa_x \pa_\xi}}
\def\mexx{e^{-\pa_x \pa_\xi}}
\def\SLinf{SL (\infty ; \IR )}             
\def\slinf{sl (\infty ; \IR )}               
\def\sumlm{\sum_{l=1}^{\infty} \sum_{\v m\v \leq l}}
\def\WDO#1{W_{DOP (S^1 )} \lb #1\rb}               


\def\PsDAS{\Psi{\cal DO} (S^1 )}
\def\PsDA{\Psi{\cal DO}}
\def\ePsDA{{\widetilde {\Psi{\cal DO}}}}
\def\dPsDA{\Psi{\cal DO}^{\ast}}
\def\PsDOS{\Psi {\rm DO} (S^1 )}
\def\PsDO{\Psi {\rm DO}}
\def\ePsDO{{\widetilde {\Psi {\rm DO}}}}
\def\Volt{\Bigl( \Psi{\cal DO} \Bigr)_{-}}
\def\eVolt{\Bigl( {\widetilde {\Psi{\cal DO}}} \Bigr)_{-}}
\def\dVolt{\Bigl( \Psi{\cal DO}^{\ast} \Bigr)_{-}}
\def\VOLT{\Bigl( \Psi {\rm DO} \Bigr)_{-}}
\def\eVOLT{\Bigl( {\widetilde {\Psi {\rm DO}}} \Bigr)_{-}}


\def\Rm#1#2{r(\vec{#1},\vec{#2})}          
\def\OR#1{{\cal O}(R_{#1})}           
\def\ORti{{\cal O}({\widetilde R})}           
\def\AdR#1{Ad_{R_{#1}}}              
\def\dAdR#1{Ad_{R_{#1}^{\ast}}}      
\def\adR#1{ad_{R_{#1}^{\ast}}}       
\def\AdB#1{Ad\Bigl( g^{-1}_{#1}(L) \Bigr)}  
\def\dAdB#1{Ad^{\ast}\Bigl( g_{#1}(L) \Bigr)}  
\def\KP{${\bf \, KP\,}$}                 
\def\KPl{${\bf \,KP_{\ell}\,}$}         
\def\KPo{${\bf \,KP_{\ell = 0}\,}$}         
\def\mKPa{${\bf \,KP_{\ell = 1}\,}$}    
\def\mKPb{${\bf \,KP_{\ell = 2}\,}$}    
\def\bKP{${\bf \, KP\,}$}                 
\def\bKPl{${\bf \,KP_{\ell}\,}$}         
\def\bKPo{${\bf \,KP_{\ell = 0}\,}$}         
\def\bmKPa{${\bf \,KP_{\ell = 1}\,}$}    
\def\bmKPb{${\bf \,KP_{\ell = 2}\,}$}    

\def\jc{J^C}
\def\bj{{\bar J}}
\def\sj{{\jmath}}
\def\bsj{{\bar \jmath}}
\def\bp{{\bar \p}}
\def\faa{Fa\'a di Bruno~}

\def\AM#1{A^{(M)}_{#1}}
\def\BM#1{B^{(M)}_{#1}}
\def\Xb{X(b_{M})}
\def\Yb{Y(b_{M})}
\def\Xbo{X_{(0)}(b_{M})}
\def\Ybo{Y_{(0)}(b_{M})}

\def\KP3{{\bf KP_{2+1}}}
\def\mKP3{{\bf mKP_{2+1}}}
\def\KPm{{\bf (m)KP_{2+1}}}
\def\KPt{{\bf KP_{1+1}}}
\def\mKPt{{\bf mKP_{1+1}}}

\def\bb{{\bar  B}}
\def\bom{{\bar \om}}
\newcommand\ttmat[9]{\left(\begin{array}{ccc}  
{#1} & {#2} & {#3} \\ {#4} & {#5} & {#6} \\
{#7} & {#8} & {#9} \end{array} \right)}
\newcommand\thrcol[3]{\left(\begin{array}{c}  
{#1} \\ {#2} \\ {#3} \end{array} \right)}

\def\cKP{{\sf cKP}~}
\def\scKP{{\sf scKP}~}
\newcommand\Back{{B\"{a}cklund}~}
\newcommand\DB{{Darboux-B\"{a}cklund}~}
\def\BH{{Burgers-Hopf}~}
\def\tQ{{\widetilde Q}}
\def\tit{{\tilde t}}
\def\hQ{{\widehat Q}}
\def\hb{{\widehat b}}
\def\hR{{\widehat R}}
\def\htt{{\hat t}}

\def\ot#1{{#1}\otimes \one + \one \otimes {#1}}
\def\ad{ad_{\ast}}                
\def\dad{ad_{\ast}^{\ast}}        
\def\ddoubl{{\cal D}^{\ast}}      
\def\edoub{{\wti {\cal D}}}             
\def\eddoub{{\wti {\cal D}^{\ast}}}     


\def\AA{\widetilde {\cal A}}
\def\ruc#1{r_{#1} (U_0 ,c)}                      

\newcommand{\nit}{\noindent}
\newcommand{\ct}[1]{\cite{#1}}
\newcommand{\bi}[1]{\bibitem{#1}}
%
%
\newcommand\PRL[3]{{\sl Phys. Rev. Lett.} {\bf#1} (#2) #3}
\newcommand\NPB[3]{{\sl Nucl. Phys.} {\bf B#1} (#2) #3}
\newcommand\NPBFS[4]{{\sl Nucl. Phys.} {\bf B#2} [FS#1] (#3) #4}
\newcommand\CMP[3]{{\sl Commun. Math. Phys.} {\bf #1} (#2) #3}
\newcommand\PRD[3]{{\sl Phys. Rev.} {\bf D#1} (#2) #3}
\newcommand\PLA[3]{{\sl Phys. Lett.} {\bf #1A} (#2) #3}
\newcommand\PLB[3]{{\sl Phys. Lett.} {\bf #1B} (#2) #3}
\newcommand\JMP[3]{{\sl J. Math. Phys.} {\bf #1} (#2) #3}
\newcommand\PTP[3]{{\sl Prog. Theor. Phys.} {\bf #1} (#2) #3}
\newcommand\SPTP[3]{{\sl Suppl. Prog. Theor. Phys.} {\bf #1} (#2) #3}
\newcommand\AoP[3]{{\sl Ann. of Phys.} {\bf #1} (#2) #3}
\newcommand\RMP[3]{{\sl Rev. Mod. Phys.} {\bf #1} (#2) #3}
\newcommand\PR[3]{{\sl Phys. Reports} {\bf #1} (#2) #3}
\newcommand\FAP[3]{{\sl Funkt. Anal. Prilozheniya} {\bf #1} (#2) #3}
\newcommand\FAaIA[3]{{\sl Functional Analysis and Its Application} {\bf #1}
(#2) #3}
\def\TAMS#1#2#3{{\sl Trans. Am. Math. Soc.} {\bf #1} (#2) #3}
\def\InvM#1#2#3{{\sl Invent. Math.} {\bf #1} (#2) #3}
\def\AdM#1#2#3{{\sl Advances in Math.} {\bf #1} (#2) #3}
\def\PNAS#1#2#3{{\sl Proc. Natl. Acad. Sci. USA} {\bf #1} (#2) #3}
\newcommand\LMP[3]{{\sl Letters in Math. Phys.} {\bf #1} (#2) #3}
\newcommand\IJMPA[3]{{\sl Int. J. Mod. Phys.} {\bf A#1} (#2) #3}
\newcommand\TMP[3]{{\sl Theor. Mat. Phys.} {\bf #1} (#2) #3}
\newcommand\JPA[3]{{\sl J. Physics} {\bf A#1} (#2) #3}
\newcommand\JSM[3]{{\sl J. Soviet Math.} {\bf #1} (#2) #3}
\newcommand\MPLA[3]{{\sl Mod. Phys. Lett.} {\bf A#1} (#2) #3}
\newcommand\JETP[3]{{\sl Sov. Phys. JETP} {\bf #1} (#2) #3}
\newcommand\JETPL[3]{{\sl  Sov. Phys. JETP Lett.} {\bf #1} (#2) #3}
\newcommand\PHSA[3]{{\sl Physica} {\bf A#1} (#2) #3}
\newcommand\PHSD[3]{{\sl Physica} {\bf D#1} (#2) #3}
\newcommand\JPSJ[3]{{\sl J. Phys. Soc. Jpn.} {\bf #1} (#2) #3}

\newcommand\hepth[1]{{\sl hep-th/#1}}

\section{Introduction}

We consider a generalized Drinfeld-Sokolov hierarchy (\cite{ds,GIH1,GIH2}),
 based on a 
Kac-Moody algebra ${\hat \lie}= {\hat {sl}} (M+ K+1)$ and 
defined by the following matrix eigenvalue problem:
\be
L \Psi = 0 \quad ;\quad L \equiv (D - A  -  E ) \quad; \quad
D \equiv I \partder{}{x}
\lab{lpsi}
\ee
The constant $E$ is a non-regular and semisimple element of ${\hat \lie}$ 
given by
\be
E = \sum_{a=1}^{K}  E^{(0)}_{\a_{M+a}}
+   E^{(1)}_{-(\a_{M+1}+ \cdots+\a_{M+K})}
\lab{ele}
\ee
The potential $A$ in \rf{lpsi} contains the dynamical variables of the
model, namely  $q_i$, $r_i$, $U_a$ and $\nu$, and is given by
(\ct{AFGZ,mcintosh}):
\be
A  = \sum_{i=1}^{M} \( q_i P_i + r_i {P}_{-i} \) +
\sum_{a=1}^{K} U_{M+a} ( \a_{M+a} \cdot H^{(0)}) + \nu \, {\hat c}
\lab{a20}
\ee
where
$P_{\pm i} = E_{\pm (\a_{i} + \a_{i+1} + \ldots +\a_{M})}^{(0)}\, ,
\,\, i=1, 2, \ldots, M$,
and ${\hat c}$ is a central element of ${\hat \lie}$. We are using the
Cartan-Weyl basis for ${\hat \lie}$ with generators $H_i^{(n)}$
and $E_{\a}^{(n)}$, with $i=1,2, \ldots {\rm rank}\, {\hat \lie}$, 
$n \in \IZ$ being the eigenvalues of the standard derivation $d$ of 
${\hat \lie}$, and  $\a$ being  
roots of the finite algebra $sl(M+ K+1)$, associated to the
affine Kac-Moody algebra ${\hat \lie}= {\hat {sl}} (M+ K+1)$. 

The integral gradations of ${\hat \lie}$ (\cite{kac}) play an important role 
in the construction of integrable hierarchies (\cite{GIH1}). 
They  are labeled by  sets of 
${\rm rank}\,{\hat \lie} + 1$ co-prime non negative integers 
$\bfs = ( s_0, s_1, \ldots, s_r)$. 
The corresponding grading operators, in the case of 
${\hat {sl}} (M+ K+1)$, are given by 
\be
Q_{{\bf s}} \equiv \sum_{a=1}^{M+ K} s_a \l_a  \cdot H^{(0)} + 
 d \, \sum_{i=0}^{M+ K} s_i
\ee
where $\l_a$ are the fundamental weights of $sl(M+ K+1)$. The relevant
gradation to the hierarchy associated to \rf{lpsi} is
\be
\bfs_{\rm cKP} = ( 1, \underbrace{0, \ldots ,0}_{M}, 
\underbrace{1, \ldots,1}_{K}\, ) 
\; ; \quad \leftrightarrow \quad 
Q_{{\rm cKP}} \equiv \sum_{a=1}^{K}  \l_{M+a} \cdot H^{(0)} + (K+1) d
\lab{svector}
\ee
Notice that $E$ and $A$ have grades $1$ and $0$ respectively, w.r.t. 
\rf{svector}. 

The gradation \rf{svector} interpolates
between the principal $ \bfs_{\rm principal} = (1,\ldots , 1)$
and the homogeneous one $ \bfs_{\rm homogeneous} = (1,0 \ldots , 0)$.
As it is well-known these two limits define   
the KdV (\cite{GIH1}) and AKNS hierarchies (\cite{FK83,AGZ}), respectively 
with 
\be
E_{\rm KdV} = E^{(1)}_{-(\a_{1}+ \cdots+\a_{K})} +   
\sum_{a=1}^{K}  E^{(0)}_{\a_{a}}
\lab{elepr}
\ee
and 
\be
E_{\rm AKNS} =  \lambda_M H^{(1)}
\lab{elehom}
\ee

The fact that $E$ is semisimple means it decomposes ${\hat \lie}$ into the
kernel and image of its adjoint action, i.e. ${\hat \lie} = \kere + \ime$. Then
one can gauge transform $L$ into $\kere$, $L \ra L_0 \equiv U \, L \, U^{-1}$, 
with $U$ being an exponentiation of generators of negative 
$\bfs_{\rm cKP}$-grade of 
${\hat \lie}$. One introduces a flow equation for each element $b^{(N)}$, in
the center of $\kere$, with positive $\bfs_{\rm cKP}$-grade $N$, 
as follows
\be
{d\, L \o {d\, t_{N}}}  ={d\, A \o {d\, t_{N}}} \equiv
\lb L \, , \, B_{N} \rb 
\lab{flow}
\ee
where $B_{N} \equiv \( U^{-1}\, b^{(N)}\, U \)_{\bfs_{\rm cKP} \geq 0}$. 
We shall choose $b^{(1)}\equiv E$, $B_1 \equiv E+A$ and $t_1\equiv x$. 

We are interested in discussing the soliton solutions for the model defined by
equation \rf{lpsi}, following the method of (\cite{luiz}). According to 
(\cite{luiz}) the basic ingredient for the appearance of soliton solutions 
is that there must exist one or several ``vacuum solutions'', 
such that the Lax operators, when evaluated on them, should lie in 
some abelian subalgebra of ${\hat \lie}$, and in addition the corresponding 
components of it should be constant. Accordingly, one requires
\be
B_N^{({\rm vac.})} \equiv \varepsilon_N = \sum_{i=0}^N c_N^i \, a_i \; ; \qquad 
[a_i,a_j] = i \>\beta_i \> \delta_{i+j,0}
\ee
where, $\beta_i$ is some constant, and the oscillators $a_i$ have 
$\bfs_{\rm cKP}$-grade $i$. Then one can write
\be
B_N^{({\rm vac.})} = {\partial \Psi^{({\rm vac.})}\over \partial t_{N}}\>
{\Psi^{-1}}^{({\rm vac.})}\> \; ; \qquad {\rm with} \qquad 
\Psi^{({\rm vac.})} = \exp \( \sum_{N} \varepsilon_N \, t_N \) 
\ee
The soliton solutions are constructed using the dressing transformation method
as follows. Choose a constant element $h$ which is an exponentiation of the
generators of ${\hat \lie}$, and perform the generalized Gauss decomposition
\br
\Psi^{({\rm vac.})}\> h\> {\Psi^{-1}}^{({\rm vac.})} &=& 
(\Psi^{({\rm vac.})}\> h\> {\Psi^{-1}}^{({\rm vac.})})_{\bfs_{\rm cKP}<0}\> 
(\Psi^{({\rm vac.})}\> h\>{\Psi^{-1}}^{({\rm vac.})})_{\bfs_{\rm cKP}=0}\> 
\times \nonu \\
&\times&
(\Psi^{({\rm vac.})}\> h\> {\Psi^{-1}}^{({\rm vac.})})_{\bfs_{\rm cKP}>0}\>.
\er
Define the group element
\br
\Psi^{h} \> & \equiv &\>  
\( (\Psi^{({\rm vac.})}\> h\> 
{\Psi^{-1}}^{({\rm vac.})})_{\bfs_{\rm cKP}<0}\>\)^{-1} \, 
\Psi^{({\rm vac.})} \, h 
\\
&  = &\>  
(\Psi^{({\rm vac.})}\> h\>{\Psi^{-1}}^{({\rm vac.})})_{\bfs_{\rm cKP}=0}\> 
(\Psi^{({\rm vac.})}\> h\> {\Psi^{-1}}^{({\rm vac.})})_{\bfs_{\rm cKP}>0}\>
\Psi^{({\rm vac.})} \nonu 
\er
Then, one can easily verify that
\be
B^h_N = {\partial \Psi^{h}\over \partial t_{N}}\>
{\Psi^{h}}^{-1}\>
\ee
has the same grading structure as $B_N$. Since $B^h_N$ is a flat connection it
automatically satisfies the Lax (or flow) equation, and therefore by equating 
$B_N$ to $B^h_N$ one gets a solution for the hierarchy for each choice of $h$.

The soliton solutions correspond to those $h$'s which are  products of
exponentials of eigenvectors of the oscillators $a_i$'s, i.e.
\be
h = e^{F_1} \> e^{F_2} \> \cdots e^{F_n} \>, \qquad 
[ \varepsilon_N \> , \> F_k ] = \omega_N^{(k)} \> F_k \> , \quad k=1,2, 
\ldots, n\>.
\lab{eigenb}
\ee
Notice, that the $F_k$'s do not have to be eigenvectors of all 
$\varepsilon_N$'s. The soliton is a solution of a given flow equation, 
associated to a time $t_N$, only if the $F_k$'s are eigenvectors of the 
corresponding $\varepsilon_N$ and also $\varepsilon_1$.

We also define, following (\cite{luiz}), the generalized Hirota tau-function as 
\br
\tau_{\mu_0,\mu_{0}'}(t)\> &=&
\> \bra{\mu_{0}'}\Psi^{({\rm vac})}\> h\> {\Psi^{({\rm
vac})}}^{-1}\ket{\mu_{0}} \nonu\\  
&=& \> \bra{\mu_{0}'}\> e^{\sum_{N} \varepsilon_N t_N}
\>  h \>  e^{-\sum_{N} \varepsilon_N t_N} \> \ket{\mu_{0}}\>, 
\lab{TauB}
\er
where $\ket{\mu_{0}}$ and $\ket{\mu_{0}^{\pr}}$ are suitably chosen 
states in an integrable
highest weight representation of ${\hat \lie}$, which are annihilated by all
generators with positive $\bfs_{\rm cKP}$-grade. 

The truncation of the Hirota's formal expansion is now understood from the
nilpotency of the operators $F_k$'s. Such property is most easily verified
using the vertex operator construction of the integrable highest weight
representation of ${\hat \lie}$. In this paper, we give the explicit
construction of the vertex operator representation relevant for the soliton
solutions of the hierarchy \rf{lpsi}. 

For pedagogical reasons, 
in section 2 we review the construction for the pure homogeneous and pure
principal gradations corresponding to untwisted and twisted vertex operators
related to AKNS and m-KDV hierarchies respectively.

In section 3 we present, following the same line of thought, the vertex functions
construction for the intermediate gradation  hierarchy.  In section 4 we
present the vertex operator algebra in terms of OPE (Operator Product
Expansion) and show that it, in fact, reproduces the original $SL(M+K+1)$
Kac-Moody algebra.  In section 5 we specialize for $K=M=1$ and  discuss the
resulting tau functions (\cite{afgz2}).

\section{Vertex Operator Construction for the Homogeneous and Principal
gradations}

Let us first consider the general Kac-Moody algebra in the Cartan-Weyl basis
\br
\lb H^m_i \, , \, H^n_j \rb &=& m\,\d_{m+n,0}\,\d_{i,j} 
\qquad i,j = 1, \ldots , \mbox{\rm rank $\lie$}
\lab{211}\\
\lb H^{m}_{i} \, , \, E^n_{\a} \rb &=& (\a )^{i} E_{\a}^{m+n}
\lab{212}\\
\lb E^m_{\a} \, ,\, E^n_{\b} \rb  &=& \left\{ 
\begin{array}{ll}
\epsilon (\a , \b ) E_{\a + \b }^{m+n} 
& \, \mbox{\rm if $\a + \b$ is a root} \\
  H_{\a}^{m+n} & \,\mbox{if $\a + \b = 0$} \ \\
  0 & \, {\rm otherwise}
  \end{array} \right. 
\lab{213}
\er
where $(H_i^m )^{\dagger} = H_i^{-m}$, $ (E_{\a}^m )^{\dagger} = E_{\a}^{-m}$. 
{}From eqns (\rf{211}, \rf{212} and \rf{213}) we will  define 
an affine Heisenberg subalgebra satisfying 
\be
[{\cal A}^m_i , ({\cal A}^n_j)^{\dagger} ] = m\, \d_{m,n} \,\d_{i,j}\quad
i,j=1,\ldots , rank \;\;\lie
\lab{heisen}
\ee

\subsection{Homogeneous Gradation}

We associate $ {\cal A}^m_i = H^m_i$ and  $ ({\cal A}^m_i)^{\dagger} =
(H^m_i)^{\dagger}$.  Define now the rank $\lie $ dimensional Fubini-Veneziano
field (see (\ct{goddardetal}))
\br
Q^i(z) &=& i \sum_{n > 0} {{\cal A}^n_i z^{-n} \o n} \nonu \\
(Q^i)^{\dagger}(z) &=& 
-i \sum_{n > 0} {({\cal A}^n_i)^{\dagger} z^{n} \o n} \nonu \\
Q^i_0(z) &=& q^i -ip^ilnz
\lab{fv}
\er
where $[q^i , p^j ] = i \d_{i,j}$ and $ p^i = H_i^0$.

The construction of step operators satisfying \rf{212} and \rf{213} is
provided by the vertex operator 
\be
V^{\a}(z) =e^{i\a Q^{\dagger}(z)}  e^{i\a Q_{0}} e^{i\a Q(z)}= {z}^{\a^2 \o 2}
 e^{i\a Q^{\dagger}(z)}e^{i\a q} e^{\a p lnz} 
 e^{i\a Q(z)}  
\lab{215}
\ee
where the vector $\a$ denotes a root of the algebra $\lie $.  It follows that 
\be
H^i(z_1) V^{\a}(z_2) = :H^i(z_1) V^{\a}(z_2) : + {(\a )^iV^{\a}(z_1) \o {z_1 -
z_2 }}
\lab{216}
\ee
Eqn. (\rf{213}) is equivalent to the product 
\be
V^{\a}(z_1) V^{\b}(z_2) = :V^{\a}(z_1) V^{\b}(z_2): 
{z_1}^{\a^2 \o 2}{z_2}^{\b^2 \o 2}(z_1 - z_2 )^{\a \b}
\lab{217}
\ee
where $: \ldots :$ denotes normal ordering in the sense 
that ${\cal A}_n$ are moved
to the right of ${{\cal A}^{\dagger}}_n$ and $p$ to the right of $q$  
for $\a , \b $ roots of $\lie $.  

\subsection{Principal Gradation }

In this case the affine Heisenberg subalgebra is defined by 
\be 
{\cal A}^a_{a+n(K+1)} = \sum _{i=1}^{K+1-a}
E^{(n)}_{\a _{i}+ \a _{i+1} + \cdots +\a _{i+a-1}}  
 + \sum _{i=1}^{a} E^{(n+1)}_{-(\a _{i}+ \a _{i+1} 
+ \cdots +\a _{i+K-a})} 
\lab{alpha2}
\ee 
with $a=1,2, \cdots  ,K$. 
It is straightforward to verify that 
 \be
 [{\cal A}^a_{a+m(K+1)}, {{{\cal A}^b}^{\dagger}}_{b+n(K+1)} ] 
 =(a+m(K+1))\d_{m,n}
 \lab{222}
 \ee
For instance, take  $g = SL(2) $ with
\be
{\cal A}^1 _{2n+1} = E_{\a}^{(n)} + E_{-\a} ^{(n+1)} \,.
\ee
Equation \rf{222} is then easily verified.  

We now define the rank $\lie $ dimensional Fubini-Veneziano field
\br
Q^{a}& = &i \sum_{n=0}^{\infty}{ {\cal A}^{a}_{a+n(K+1)} z^{-a-n(K+1)} \o
{a+n(k+1)}} \nonu \\
{Q^{a}}^{\dagger}& = &-i \sum_{n=0}^{\infty}{ {{{\cal
A}^{a}}^{\dagger}_{a+n(K+1)}} z^{a+n(K+1)} \o
{a+n(K+1)}}
\lab{FBTwis}
\er
Notice that in this case the zero modes are absent.

We now seek within the Kac-Moody algebra ${\hat {sl}}(K+1)$ 
the eigenvalues and eigenfunctions of the affine  Heisenberg subalgebra  
defined in \rf{222}.  
We therefore find that
\be   
\sbr{{\cal A}^a_{a+n(K+1)}}{ F_{b,l} (z)} =  \omega^{- al } 
\( \omega^{ab} - 1 \)  z^{a+(K+1)n} F_{b,l} (z)     
\lab{eigen} 
\ee
with $a,b=1,\ldots ,K \; , \; l=1,\ldots , K ,K+1$, and 
where $\om = exp{{2\pi i \o {K+1}}}$,  $F_{b,l}$ is a linear combination of 
Kac-Moody generators of ${\hat {sl}}(K+1)$ and
will be given explicitly in the next section for a more general case.
The eigenvalues in equation \rf{eigenb} define the roots of $sl(K+1)$.  Our
task is now to find a set of simple roots lying in a $K$ dimensional complex
space such that any root defined in
\rf{eigenb} may be written as a integer linear combination with all
coefficients being positive (or negative).  The following set of simple roots
satisfy our requirements:
\be
 {\a}_{(K)_i} = \( \om^{i-1}(\om - 1), \om^{2(i-1)}(\om^2 - 1), 
\ldots ,\om^{K(i-1)} (\om^K - 1) \) 
\lab{sroots}
\ee
for $i=1,\ldots ,K$.  
Define now the vertex operator satisfying the eigenvalue equation \rf{eigenb}
with eigenvalue $\a$ to be
\be
V^{\a}(z) = e^{i\a^{*} Q(z)^{\dagger} } e^{i \a Q(z)}
\lab{twisver}
\ee
The OPE of vertices \rf{twisver} is a straightforward calculations yielding 
\be
V^{\a}(z_1) V^{\b}(z_2) = :V^{\a}(z_1) V^{\b}(z_2): \prod_{p=1}^{K+1} 
(1-{z_2\om ^{-p} \o z_1})^{\sum_{c=1}^{k} {\a^{c} {\b^{c}}^{*} 
\om ^{pc} \o {K+1}}}
\lab{prodver2}
\ee
Since the simple roots in \rf{sroots} satisfy the same addition table as the
simple roots of $sl(K+1)$ the product of vertices \rf{prodver} closes into
the OPE algebra of ${\hat {sl}}(K+1)$.
\section{Vertex Construction for the Intermediate Gradation}

In this section we shall discuss the vertex functions for the model defined by
eqn. \rf{lpsi} with the constant element $E$ lying in the Kac-Moody algebra
$sl(M+K+1)$ given by \rf{ele} and gradation given by \rf{svector}.  
Consider the Heisenberg algebra associated to the center of $\kere$ 
and consisting of :

\nit
1)~ \underbar{Homogeneous part of ${\hat {sl}} (M)$}
\be 
{\cal B}_i^{(n)} = \a_i \cdot H^{(n)}  \quad, \quad i=1,2, \ldots, M-1  
\lab{beta}  
\ee 

\nit
2)~ \underbar{Principal part of ${\hat {sl}} (K+1)$}
\br  
{\cal A}^a_{a+n(K+1)} &=& \sum _{i=1}^{K+1-a}
E^{(n)}_{\a _{i+M}+ \a _{i+M+1} + \cdots +\a _{i+M+a-1}} \lab{alpha}\\  
& +& \sum _{i=1}^{a} E^{(n+1)}_{-(\a _{i+M}+ \a _{i+M+1} 
+ \cdots +\a _{i+M+K-a})}\quad , \quad a=1,2, \ldots  ,K
\nonu
\er 
\nit
3)~ \underbar{``A border part''}
\be  
{\cal A}^0_{n(K+1)} = \sqrt{M+K+1 \o M} \, \l_M . H^{(n)} \, - \,{K \o 2}
\, \sqrt{M \o M+K+1} \,\, \d_{n,0} 
\lab{alphaz}  
\ee
The above elements of the Heisenberg subalgebra enter the oscillator algebra 
relations (we put $c=1$):
\be  
\sbr{{\cal A}^a_{a+n(K+1)}}{{\cal A}^{b\, \dag}_{b+m(K+1)}}= 
\( a+(K+1)n\) \d_{n,m}\d^{a,b}  \quad;\quad a,b=1, \cdots , K
\lab{osca}  
\ee
with ${\cal A}^{a\, \dag}_{a+m(K+1)} = {\cal A}^{K-a+1}_{-a-m(K+1)}$,
\be 
\sbr{{\cal A}^0_{n(K+1)}}{{\cal A}^{0\, \dag}_{m(K+1)}}= (K+1)\,n\, \d_{n,m}  
\quad ;\quad \sbr{{\cal A}^0_n}{{\cal A}^{a\, \dag}_m}=0 
\lab{osca0}  
\ee  
and 
\be  
\sbr{{\cal B}_i^{(m)}}{{\cal B}_j^{(n)}} = K_{i,j}\, m \,\d_{m+n,0}  \quad
;\quad i,j=1,\ldots ,M-1
\lab{oscb}  
\ee  
where  $ K_{i,j} $ is the Cartan matrix of $sl (M)$.  

Instead of ${\cal B}_i^{(m)}$, it is more convenient to work with 
\be
\cK^{(n)}_i \, = \, {\sum_{p=1}^i p \,{\cal B}_p^{(n)} \o N_i}
\qquad; \quad  N_i \equiv \sqrt{i (i+1)}
\lab{knl}
\ee
such that
$\Tr \( \cK^{(n)}_i \cK^{(m)}_{j}\)= \d_{i\,j}\, \d_{m+n,0}$
and
\be  
\sbr{{\cal K}_i^{(m)}}{{\cal K}_{j}^{n}} = \, m \,\d_{m+n,0} \, 
\d_{i\,j} \quad; \quad i,j=1,\ldots ,M-1
\lab{oscbck}  
\ee 

To summarize, we had parametrized the Heisenberg subalgebra in terms of
elements:
\be
{\cal A}^a_{a+n(K+1)}\;\, , \;\, {\cal A}^0_{n(K+1)}\;\, ,\; \, 
{\cal K}_i^{(n)}\quad ; \quad a=1,2, \ldots, K \quad;\quad  i=1,2,\ldots ,M-1
\lab{heisel}
\ee
which for $n=0$ constitute a Cartan subalgebra of $sl (M+K+1)$.

We work with the eigenstates of this subalgebra having the following form
\br  
F_{a,l} &=& {{\hat c}\o {(w^a -1)}} 
+ \sum_{n\in Z} z^{-n(K+1)} \sum_{i=1}^{K}
H_{M+i}^{(n)} \sum_{p=1}^{i} w^{a(i-p)} \nonu \\
&+& \sum_{b=1}^{K}
\sum_{n\in Z} w^{bl} z^{-(b+n(K+1))}\(\sum_{i=1}^{K+1-b} w^{a(i-1)}
E^{(n)}_{\a_{i+M} +\a_{i+M+1} + \cdots +\a_{i+M+b-1}}\right. \nonu \\  
& +& \left. \sum_{i=1}^{b} w^{a(i+K-b)}E^{(n+1)}_{-(\a _{i+M} + \a _{i+M+1}
+ \cdots + \a_{i+M+K-b})}\) \lab{fal}   
\nonu
\er
where $a=1,\ldots ,K \; , \; l=1,\ldots , K ,K+1 $
and $\omega$ is a non-trivial ${K+1}$-th root of unity ($\omega^{K+1} =1$).
\be  
{\bar F}_{r,l} = \sum_{n\in \IZ} z^{-n(K+1)} \sum_{p=0}^{K}  
w^{pl} z^{-p} E^{(n)}_{\a_{r} + \cdots + \a_{M+p}},  
\lab{bfrl}
\ee  
with $r=1, \ldots, M \; \;, \quad l= 1,\ldots , K ,K+1$, and 
\be  
F_{\a_j} = \sum_{n \in \IZ} z^{-n} E^{(n)}_{\a_j} \quad ;\quad j=1,\ldots ,M-1 
\lab{faj}
\ee  
One finds the following form of algebraic relations for the eigenstates 
of the Heisenberg subalgebra:
\br  
\sbr{{\cal A}^a_{a+n(K+1)}}{ F_{b,l} (z)} &=&  \omega^{- al } 
\( \omega^{ab} - 1 \)  z^{a+(K+1)n} F_{b,l} (z)  
   \nonu \\  
\sbr{{\cal A}^0_{n(K+1)} }{ F_{b,l} (z)}&=& 0 \nonu \\  
\sbr{{\cal K}_i^{(n)} }{ F_{b,l} (z)}&=& 0 \lab{aanfjz}
\er
where $a,b=1,\ldots ,K$ and $ l=1,\ldots , K ,K+1$, 
\br
\sbr{{\cal A}^a_{a+n(K+1)}}{{\bar F}_{r,l} (z)}&=& \omega^{-al} z^{a+ n(K+1)} 
{\bar F}_{r,l} (z) \nonu \\  
\sbr{{\cal A}^0_{n(K+1)} }{{\bar F}_{r,l} (z)}&=&  \sqrt{{K+1 \o \l_M^2}} 
z^{n(K+1)}\,  {\bar F}_{r,l} (z) \nonu \\  
\sbr{{\cal K}^{(n)}_i}{{\bar F}_{r,l} (z)}&=& {1 \o N_i} \(  \sum_{p=1}^i
\d_{r,p}  - \d_{r,i+1}\) z^{n(K+1)}\,  {\bar F}_{r,l} (z)
\lab{biafjl} 
\er
where $N_i = \sqrt{i (i+1)}$, $r=1,\ldots ,M$ and  $l=1,\ldots , K,K+1$.  

Furthermore, we also have
\br  
\sbr{{\cal A}^a_{a+n(K+1)}}{ F_{\a_j} (z)} &=&  0   \nonu \\  
\sbr{{\cal A}^0_{n(K+1)} }{ F_{\a_j} (z)}&=& 0 \lab{lmhfd}  \\  
\sbr{{\cal K}_i^{(n)} }{ F_{\a_j} (z)} &=&  
{1 \o N_i} \(  \sum_{p=1}^i 2 p \d_{p,j}  - p \d_{p,j-1}  - p \d_{p,j+1}\) 
z^{n}\,  F_{\a_j} (z) 
\nonu
\er
where $i,j=1,\ldots , M-1$. 

Since, $F_{j,l}, {\bar F}_{r,l} $ and $F_{\a_j}$ are step operators
associated with the Cartan subalgebra defined by the Heisenberg subalgebra
they correspond to the roots of $sl (M+K+1)$.
The simple roots defining these steps operators are:
\br 
\a_i& =& \( \a_{(M)_i}, 0, 0_{(K)} \), \quad i= 1, \ldots, M-1\nonu \\
 &=& \(   0_{(M-2)}, {{1-M }\o {N_{M-1}}}, 
\sqrt{K+1 \o \l_M^2},1_{(K)} \) \quad  for \;\;i=M \nonu \\
&=& \(0_{(M-1)}, 0, \a_{(K)_j} \), \quad for \;\; i=M+j, \;\; j=1, \ldots, K
\er
where  $ 0_{(N)} $ denotes the N-dimensional null vector, $1_{(N)}$ is the
N-dimensional vector with N unit components. The root vectors $  \a_{(M)_i}$
are 
\br
\a_{(M)_1} &= &\({2\o N_{1}}, 0_{(M-2)} \)\nonu \\
\a_{(M)_2} &= &\({-1 \o N_1},{3\o N_{2}}, 0_{(M-3)} \)\nonu \\
\vdots & \vdots & \vdots \nonu\\
\a_{(M)_i} &= &\(0_{(i-2)},-{(i-1)\o N_{i-1}}, {(i+1) \o N_i}, 0_{(M-1-i)}\), 
i=3,\ldots, M-1 
\lab{a12}
\er
and $\a_{(K)_i}$ are given in eqn. \rf{sroots}.

The step operators associated to the simple roots of \rf{a12}, \rf{sroots}
are nothing but the eigenstates of the Heisenberg algebra \rf{heisel}
and the association is 
\br
E_{{ \a}_i} &\leftrightarrow& F_{\a_i}\;\; ( = E_{\a_i}~{\rm of}~sl(M)) 
\quad;\quad i=1,2,\ldots, M-1   \lab{slm}\\
E_{{ \a}_M} &\leftrightarrow& {\bar F}_{M,K+1}  \lab{slma}\\
E_{{ \a}_{M+a}} &\leftrightarrow& F_{1, K-a+2}   
\quad ;\quad a=1,2, \ldots ,K
\lab{slmb}
\er
The remaining eigenstates are associated to the general roots of the form
${ \a} = { \a}_i + \ldots + { \a}_j$.
For instance $F_{2,0} \leftrightarrow E_{{ \a}_{M+1}+{ \a}_{M+2}}$
with
${ \a}_{M+1}+{ \a}_{M+2} = \( 0, \ldots , 0,0, \om^2 - 1, \ldots
,\om^{2K} - 1 \)$.

Define now the Fubini-Veneziano operators
\br
Q^i_0 (z) &=& q^i -i p^i\,\ln z \; ; \; Q^i(z) = i \sumi{n=1}
{ \cK_i^{(n)} z^{-n} \o n} \; ;\; i=1,\ldots, M-1 \phantom{aa}\lab{qi}\\
Q^M_0 (z) &=& q^M -i p^M\,\ln z; \quad Q^M(z) 
= i \sumi{n=1} { {\cal A}^0_{n(K+1)} z^{-n(K+1)} 
\o n(K+1)}  \lab{qm}\\
Q^{M+a} (z) &=& i \sumi{n=0} { {\cal A}^a_{a+n(K+1)} z^{a+n(K+1)} 
  \o a+ n(K+1)} 
\quad ;\quad a=1,2,\ldots, K  \lab{qma}
\er
where $p_M$ is equal to ${\cal A}^0_{n=0}$ from expression \rf{alphaz}
and the zero modes satisfy $\sbr{p^i}{q^j}= - i \d^{ij}$.

The corresponding conjugated Fu\-bi\-ni-Ve\-ne\-zia\-no ope\-ra\-tors 
$Q^{\dag} (z)$ are
obtained from \rf{qi}-\rf{qma} by taking into consideration rules
${ \cK_i^{(n)}}^{\dag} = \cK_i^{(-n)}$, $\({\cal A}^0_{n(K+1)}\)^{\dag}=
{\cal A}^0_{-n(K+1)}$, $\({\cal A}^a_{a+n(K+1)}\)^{\dag} =
{\cal A}^a_{a-n(K+1)}$ as well as $z^{\dag} = z^{-1}$.

The total number of $Q$'s equals $M+K$ which is the rank of $sl (M+K+1)$.

Putting together the simple root structure from 
eqs.\rf{a12}, \rf{sroots} with the Fubini-Veneziano operators
enables us to write down a compact expression for the general vertex operator
in the normal ordered form:
\br 
V^{\a} (z)\; &\equiv& \;z^{\h  (\a_{(M)})^2}\, \times \lab{vertex}\\
&\times&
\exp \({i {\vec \a}^{\ast} \cdot {\vec Q}^{\dag} (z)}\)\,
\exp \({i {\vec \a} \cdot {\vec q}}\)\, \exp \({{\vec \a} \cdot {\vec p} \ln
z}\) \, \exp \({i {\vec \a}\cdot {\vec Q} (z)}\)
\nonu
\er
where $(\a_{(M)})^2 = \sum_{j=1}^M (\a_{(M)}^j)^2$ i.e. the sum of squares of
components in the $M-1$ subspace, and
\br
{\vec Q} &=& \( Q_1, Q_2, \ldots ,Q_M,Q_{M+1}, \ldots, Q_{M+K} \) \lab{vecq}\\
{\vec q} &=& \( q_1, q_2, \ldots ,q_M,{0}_{(K)}
\) \lab{vecsq}\\
{\vec p} &=& \( p_1, p_2, \ldots ,p_M,{0}_{(K)}
\) \lab{vecsp}
\er

The product of two vertex operators is found to be:
\br
&& V^{\a} (z_1) V^{\b} (z_2) = : V^{\a} (z_1) V^{\b} (z_2):
z_1^{\h  \sum_{j=1}^M \( (\a^j)^2+2 \a^j \b^j \)}
z_2^{\h  \sum_{j=1}^M  \b^j \b^j} \nonu\\
&\times & \( 1 - {z_2 \o z_1}\)^{\sum_{j=1}^{M-1} \( \a^j \b^j + 
 (\a^M \b^M /(K+1)) +\sum_{a=1}^{K} (\a^{M+a} {\b^{M+a}}^{\ast} /(K+1))\)} 
\nonu \\
&\times& \prod_{p=1}^K \( 1 - {z_2 \o z_1} \om^{-p} \)^{ (\a^M \b^M /(K+1)) +
\sum_{a=1}^{K} \om^{al}(\a^{M+a} {\b^{M+a}}^{\ast} /(K+1))   }
\lab{prodver}
\er
where the $*$ stands for complex conjugation.
Similarly for the more complicated products of vertices.

It is also straightforward to see that the square of a vertex vanishes.
In order to consider the product of several vertex operators we first define
the generalized scalar product for $M+K$ component vectors 
\be
(a\odot b)_p = \d_{p,0} \sum_{i=1}^{M-1} a_i b^{*}_i + {{a_i b^{*}_i }\o {K+1}}
+ \sum _{i=1}^{K} {{a_{M+i} b^{*}_{M+i} }\o {K+1}}\om ^{ip}
\lab{odot}
\ee
With this new notation, eqn (\rf{prodver}) becomes
\be
 V^{\a} (z_1) V^{\b} (z_2) = : V^{\a} (z_1) V^{\b} (z_2):
z_1^{\h   (\a+  \b)^2 }
\({z_2 \o z_1 }\)^{\h   \b ^2 }  
 \prod_{p=0}^K \( 1 - {z_2 \o z_1} \om^{-p} \)^{ (\a \odot \b)_p } 
\lab{prodverodot}
\ee
The general formula for the product of several vertices is therefore given as 
\br
V^{\g_1}(z_1)\cdots V^{\g_N}(z_N)& =& :V^{\g_1}(z_1)\cdots V^{\g_N}(z_N) 
:\prod _{m=1}^{N-1} z_{m}^{\g_m (\g_{m+1}+\g_{m+2} + \cdots + \g_{N})} \nonu \\
&\times & \prod _{r=2}^{N}
\prod _{i<r} \prod _{p=0}^{K} \( 1 - {z_r \om ^{-p} \o z_{r-i}} \)^{\g_r \odot
\g_{r-i}} 
\lab{sevvert}
\er

\section{The OPE Algebra of Vertex Operators}

We have already proposed a set of $M+K$ simple roots satisfying the same
addition properties of those of $sl(M+K+1)$.  We now use the product of
vertices given in \rf{prodver} to obtain the OPE to show that, in fact, it
reproduces the algebra ${\hat {sl}}(M+K+1)$.  Let us associate the Kac-Moody 
currents to the vertex operators via
\be
e^{(M)}_{\a} = V^{\a}(z)
\ee
for roots in the $sl(M)$ sector, i.e. $\a =(\a_{(M)}, 0, 0_{(K )})$ where
$\a_{(M)}$ denote a root in the pure $sl(M)$ subalgebra,
\be
e^{(K)}_{\a} = {V^{\a}(z) \o |1-\om ^{\#}| }
\ee
for roots in the $sl(K+1)$ sector, i.e. $\a =(0_{(M)}, 0, \a_{(K)} )$
where $\a_{(K)}$ is a root in the $sl(K+1)$ sector and 
$\#$ denote the number of simple roots in $\a_{(K)}$, 
\be
e^{(MK)}_{\a} = {V^{\a}(z) \o  {(K+1)}^{1\o 2M} }
\ee
for roots of the form $\a =(\a_{(M)}, \a_{(MK)}, \a_{(K)} )$, where 
$\a_{(MK)}$ denote the $M$-th component of the root $\a $.

It follows from the complex structure of the simple roots given in the
previous section, that the OPE algebra  displaying the most singular part of
the product of vertices is given by
\be
V^{\a}(z_1) V^{\b}(z_2)  = \left\{ \begin{array}{ll}
{z_1 V^{\a + \b}(z_1) \o {z_1 - z_2 }} & \,{\rm  if}
\;\; \a +\b \;\; {\rm is }\;\; {\rm a }\;\; {\rm root } \\
 {z_1 \a H(z_1) \o {z_1 - z_2}} - {z_1 {d \o {dz_1}} ({z_1 \o {z_1 -
z_2}})} & \, {\rm if} \;\; \a + \b =0   \\
 0 & \, {\rm otherwise }
 \end{array}  \right.
\ee

\section{Special Case M=K=1; Tau Functions}
We consider now the special case of $M=K=1$ with 
${\hat \lie} = {\hat {sl}} (3)$ and the Lax matrix operator from \rf{lpsi}
with $A$ from \rf{a20} and $E$ from \rf{ele} being given by (see (\cite{afgz2})
\be
L = D - \ttmat{0}{q}{0}
             {r}{U_2}{1}
             {0}{\l}{-U_2}
- \nu {\hat c}	     
\lab{ttmat}
\ee
where $\l$ is the usual loop parameter.
We now describe two different two-soliton solutions obtained from the above
vertex construction.

The first one has $r=0$ and $q \ne 0$ and equal
to 
\be
q = - \sqrt{2}\,\,{z_2}\,e^{(\,{ t_3}\,{z_2}^{3} + {t}\,{z_2}^{2} + {x}\,
{z_2}\,)}\, \left(  \,1
 + \h\,{e}^{(\, - 2\,{x}\,{z_1} - 2\,{t_3}\,{ z_1}^{3}\,)}\,\({z_1+ z_2\o z_1
 -  z_2} \)  \, \!  \right) / \t^{(0)}_0
\lab{qyes}
\ee
with the tau-functions $\t^{(0)}_0$ and $\t^{(0)}_2$ 
\be
\t^{(0)}_0 =  1 - \h \, e^{(\, - 2\,{x}\,{ z_1} - 2\,{ t_3}\,
{ z_1}^{3}\,)} \quad ;\quad
\t^{(0)}_2 =  1 + \h \, e^{(\, - 2\,{x}\,{ z_1} - 2\,{ t_3}\,
{ z_1}^{3}\,)}
\lab{yes}
\ee
from which we obtain $U_2$ and $\nu$ by using:
\be
U_2  = - \pa_x \ln \( { \t^{(0)}_{0} \o \t^{(0)}_{2} }\) \qquad ;
\qquad \nu = - \pa_x \ln \(  \t^{(0)}_{0}\) \lab{u2nsol}
\ee

Another two-soliton solution for which this time
both $q \ne 0$ and $r \ne 0$ is:
\be
\t^{(0)}_{\s} = 1 + (-1)^{(\s/2)} \,2
\,{\displaystyle \frac {{ z_1}^{1+\s/2}\,{ z_2}^{2-\s/2}\,{\rm e}^{(\,{x}\,
{ z_1} + {t}\,{ z_1}^{2} +t_3 z_1^3+ {x}\,{ z_2} - {t}\,{ z_2}^{2}+t_3 z_2^3
\,)}}{(\,{ z_1} - { z_2}\,)\,(\,{ z_1} + { z_2}\,)^{2}}}
\lab{t02sat}
\ee
with $\s =0,2$, and 
\be
r = \frac{\sqrt{2}\,{ z_2}\,{\rm e}^{(\,
 - {t}\,{ z_2}^{2} + {x}\,{ z_2}+t_3 z_2^3\,)}}{\t^{(0)}_2}
\qquad ; \qquad
q=  \frac{\sqrt{2}\,{ z_1}\,{\rm e}^{(\,{t
}\,{ z_1}^{2} + {x}\,{ z_1}+t_3 z_1^3\,)}}{\t^{(0)}_0 }
\lab{rqsat}
\ee
and again $U_2$ and $\nu$ can be obtained from \rf{u2nsol}.

In the above examples we only kept the times $t_n$ with $n \leq 3$ for which
we verified validity of the relevant evolution equations.

The novel feature of the above soliton solutions
is that they mix ex\-po\-nen\-tials $\exp \( \sumi{n=1} t_n z_j^n\)$ 
which represent a typical time dependence for the KP solutions 
with pure KdV time dependence of the type 
$\exp \( \sumi{n=0} t_{2n+1} z_j^{2n+1}\)$ involving only odd times
(\cite{afgz2}).

\section{APPENDIX}

We now provide some useful relations extensively used in obtaining the
formulas given in the text:
\be
\sum_{n=0}^{\infty} {x^{a+(K+1)n} \o {a+(K+1)n}} = {- 1\o
{K+1}}\sum_{p=1}^{K+1} \om ^{ap} ln (1- x\om ^{-p}) \quad ; \; \; \; 
a=1, \ldots , K
\lab{a1}
\ee
for
\be
\om ^{K+1}=1 \enspace .
\lab{a2}
\ee
It also follows that
\be
1+ \om + \cdots + \om ^K = 0
\lab{a3}
\ee
and
\be
(1- x^{K+1}) = \prod_{p=1}^{K+1} (1- \om ^{p} x)
\lab{a4}
\ee
from where we obtain, after using L'Hopital's rule
\be
K+1 = \prod_{p=1}^{K} (1-\om ^{p})
\ee
and
\be
\sum_{p=1}^{K} {1\o {1- \om ^p }} = {K\o 2}
\ee
\lskip
{\bf Acknowledgements} A.H.Z. would like to thanks UIC and FAPESP 
for the hospitality and financial support. L.A.F., J.F.G. and A.H.Z.
acknowledge partial financial support from CNPq.  
H.A.'s work was supported in part by the U.S. Department of Energy Grant No.
DE-FG02-84ER40173.
\lskip

\end{document}